\documentclass[aip,apl,reprint,showpacs,showkeys,preprintnumbers,superscriptaddress,amsmath,amssymb,showkeys]{revtex4-2}

\usepackage{graphicx}
\usepackage{dcolumn}
\usepackage{bm}
\usepackage{hyperref}
\usepackage[dvips]{color}

\begin{document}




\title{Electrical transport properties of thick and thin Ta-doped SnO$_2$ films}


\author{Zong-Hui Gao}
\affiliation{Tianjin Key Laboratory of Low Dimensional Materials Physics and
Preparing Technology, Department of Physics, Tianjin University, Tianjin 300354,
China}
\author{Zi-Xiao Wang}
\affiliation{Tianjin Key Laboratory of Low Dimensional Materials Physics and
Preparing Technology, Department of Physics, Tianjin University, Tianjin 300354,
China}
\author{Dong-Yu Hou}
\affiliation{Tianjin Key Laboratory of Low Dimensional Materials Physics and
Preparing Technology, Department of Physics, Tianjin University, Tianjin 300354,
China}
\author{Xin-Dian Liu}
\affiliation{Tianjin Key Laboratory of Low Dimensional Materials Physics and
Preparing Technology, Department of Physics, Tianjin University, Tianjin 300354,
China}
\author{Zhi-Qing Li}
\email[Corresponding Author, Email: ]{zhiqingli@tju.edu.cn}
\affiliation{Tianjin Key Laboratory of Low Dimensional Materials Physics and
Preparing Technology, Department of Physics, Tianjin University, Tianjin 300354,
China}



\date{\today}

\begin{abstract}
Ta-doped SnO$_2$ films with high conductivity and high optical transparency have been successfully fabricated using rf-sputtering method and their electrical transport properties have been investigated. All films reveal degenerate semiconductor (metal) characteristics in electrical transport properties. For the thick films ($t\sim 1\,\mu \rm{m}$ with $t$ being the thickness) deposited in pure argon, the electron-phonon scattering alone cannot explain the temperature dependent behaviors of resistivity, the interference effect between electron-phonon and electron-impurity scattering should be considered. For the $t\lesssim 36$\,nm films, both conductivity and the Hall coefficient show linear relation with the logarithm of temperature ($\ln T$) from $\sim$100\,K down to liquid helium temperature. The $\ln T$ behaviors of conductivity and Hall coefficient cannot be explained by the Altshuler-Aronov type electron-electron interaction effect, but can be quantitatively interpreted by the electron-electron interaction effects in the presence of granularity. Our results not only provide strong supports for the theoretical results on the electron-phonon-impurity interference effect, but also confirm the validity of the theoretical predictions of charge transport in granular metals in strong coupling regime.
\end{abstract}


\maketitle


\section{Introduction}\label{SecI}
Transparent conducting oxide (TCO) films possess both high conductivity and high optical transparency in the visible range, which render them to be widely used in the fields of flat panel displays, photovoltaic electrochromics, solar cells, and energy-efficient windows.~\cite{Jarzebski1982,exarhos_discovery-based_2007,David2010,Gilshtein2020} Among various TCO films, SnO$_2$-based films have long plasma wavelength, higher thermal stability, excellent durability, and higher mechanical hardness.~\cite{Singh2008,Kim2008,Gordon2000,Hu2011,king_conductivity_2011} Thus they are the major industrial TCO films used in the fields of low-emissivity windows and solar cells. Currently, SnO$_2$-based TCO films are mainly made of fluorine-doped SnO$_2$ (FTO) and antimony-doped SnO$_2$ (ATO) materials.~\cite{gao_solid-state_2012,islam_alkaline_2020} For example, billions of square feet of FTO-coated window glass have been installed in buildings around the world. The architectural FTO films are generally deposited by spray pyrolysis method and their good conductivity is sacrificed for infrared radiation reflectivity and high visible transparency.~\cite{Gordon2000,rakhshani_electronic_1998,shi_roughness_2014} FTO or ATO film with high conductivity is usually prepared by chemical vapour deposition techniques and not suitable to be deposited by sputtering methods.~\cite{suh_atmospheric-pressure_1999,lee_structural_2006,kim_doping_2016,jain_electrical_2004,kim_transparent_2008} Thus to explore SnO$_2$-based TCO films which could be fabricated by sputtering deposition is nontrivial.

Recently, the first-principle calculation results\cite{schleife_ab_2012,darriba_ab_2014,behtash_electronic_2015,behtash_electronic_2015} indicate that Ta-doped SnO$_2$ (Ta:SnO$_2$) reveals degenerate semiconductor characteristics in electronic structure and its optical bandgap is wider than that of the undoped SnO$_2$. It means that Ta:SnO$_2$ could reveal metallic characteristics in transport property and remain visibly transparent. Using a target composed of mixture of SnO$_2$ and Ta$_2$O$_5$, one can fabricate Ta:SnO$_2$ film via rf-sputtering method.  Thus it is necessary to investigate the influences of deposition conditions on the resistivity and optical transparency of the deposited films. In addition, to understand the conduction mechanisms, it is essential to investigate the temperature dependence of the resistivity and Hall coefficient of the material. In this paper, Ta:SnO$_2$ films are successfully fabricated using rf-sputtering technique and the deposition conditions including substrate temperature and sputtering atmosphere are researched and optimized. The electrical transport properties for films with thickness ranging from $\sim$1000 to $\sim$10 nm are systematically investigated.

\section{Experimental method}\label{SecEM}
Our films were deposited on (0001) Al$_2$O$_3$ single crystal substrates by standard rf-sputtering method. Besides higher melting point,  Al$_2$O$_3$ single crystal possesses relatively higher thermal conductivity, which makes the thermal equilibrium between the film and sample holder more easy in electrical transport measurements. The sputtering source was a commercial ceramic target of SnO$_2$ (99.9\%) and Ta$_2$O$_5$ (99.99\%) mixture, in which the atomic ratio of Ta to Sn is $3:47$. The base pressure of the depositing  chamber was less than $1\times 10^{-4}$ Pa, and the deposition was done at 0.6\,Pa in a mixture of argon and oxygen atmosphere. Four films with thicknesses $\sim$1 $\mu$m were first deposited at $T_s=933$, 953, 963, and 983\,K ($T_s$ is the substrate temperature), respectively. During the deposition, the sputtering power was maintained at 150\,W and the volume ratio of oxygen to argon was set to zero. It was found that the resistivity of film deposited at 963\,K was less than that of other films at a certain measured temperature. Thus the substrate temperature is kept at 963\,K in the subsequent deposition processes. The volume ratio of oxygen to argon and oxygen ($O_{\rm pp}$) was then adjusted between 0 and 2.0\% and it was found that $O_{\rm pp}=0.5\%$ was the optimal ratio in terms of the conductivity and optical transparency of the film. Finally, Ta:SnO$_2$ films with thickness ranging from $\sim$11 to $\sim$36\,nm were deposited at the optimized condition ($T_s= 963$\,K and $O_{\rm pp}=0.5\%$).

The thicknesses of the $t>200$\,nm films were measured using a surface profiler (Dektak, 6M), while the thicknesses of the $t\lesssim 36$\,nm films were evaluated through growth rate and deposition time. The crystal structure and quality of the films were determined by x-ray diffraction (XRD, Rigaku D/MAX-2500). Optical absorption and transmittance spectra were measured in a UV-VIS-NIR  spectrometer (Hitachi V4100). The resistivity and Hall coefficient were measured in a physical property measurement system (PPMS-6000, Quantum Design) using the four-probe method. Hall-bar shaped samples, defined by mechanical masks, were used for the resistivity and Hall effect measurements.

\begin{figure}[htp]
	\includegraphics[scale=1.05]{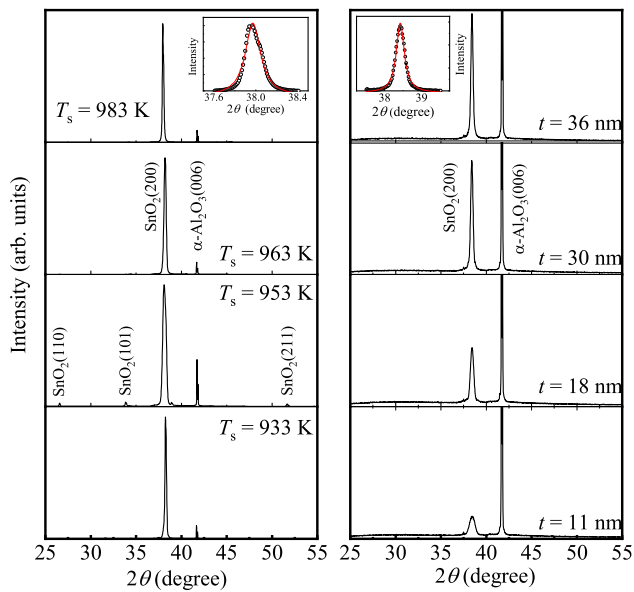}
	\caption{(Color online) Left panel: XRD patterns for the thick Ta:SnO$_2$ films deposited at different substrate temperatures $T_s$; Right panel:  XRD patterns for the thin Ta:SnO$_2$ films with different thicknesses deposited at $T_s=963$\,K.  Insets in left panel: enlarged view of the (2\,0\,0) diffraction for the thick film deposited at $T_s=983$\,K; Inset in right panel:  enlarged view of the (2\,0\,0) peak for the 36-nm-thick film deposited at $T_s=963$\,K. The solid curves in the insets are least-squares fits to Gaussian function.}\label{Fig-XRD}
\end{figure}

\section{Results and Discussion}\label{Result-Discu}
\subsection{Crystal structure and optical transparency}\label{Sub-S1}
Our samples are split into two groups. The films with thickness $t\gtrsim 550$\,nm are called thick films, while these $t\lesssim36$\,nm ones are called thin films.
Figure~\ref{Fig-XRD} shows the XRD patterns for some representative films. The deposition condition and the thickness of each film are labeled in the figure. The diffractions can be indexed based on the structure of undoped tetragonal SnO$_2$ (powder-diffraction file number: 46-1088).  From the left panel of Fig.~\ref{Fig-XRD}, one can see that the peaks related to the diffractions of (1\,1\,0), (1\,0\,1), (2\,0\,0), and (2\,1\,1) planes of tetragonal SnO$_2$ appear in the patterns for films deposited at 933\,K and 953\,K. While for the films deposited at $T_s \gtrsim963$\,K, only the diffraction of (2\,0\,0) plane can be observed. Other impurity phases such as Ta$_2$O$_5$ or Ta-Sn alloys (compounds) are not  detected. The thin films deposited at 963 K retain the texture-growth feature (along [1\,0\,0] direction), even when the thickness of the film is as low as $\sim$11\,nm.  For the thick films, it is found that the film deposited at $T_s=963$\,K has the minimum resistivity at a certain fixed temperature. Thus we take $T_s=963$\,K as the optimal substrate temperature.

The mean grain size of each film was determined from the full width at half maximum (FWHM) of (2\,0\,0) diffraction peak using Scherrer's equation\cite{birks_particle_1946,ishikawa_size_1988,yang_nanoscaled_2002,mariappan_synthesis_2006}
\begin{equation}\label{Eq-Scherrer}
d=K\lambda/(B\cos\theta)
\end{equation}
where $d$ is the mean diameter of the grains, $\lambda$ is the x-ray wavelength, $B$ is the FWHM of the selected diffraction peak, $\theta$ is the diffraction angle, and $K$ is a constant. The FWHM of (2\,0\,0) peak can be evaluated by fitting the experimental data to Gaussian function (see the inset of Fig.~\ref{Fig-XRD}). Taking $K=0.89$,~\cite{birks_particle_1946} we calculate the mean grain sizes of the films and list them in Table~\ref{Table-I}. For the thin films, the mean grain size increases as the thickness of the films increases, while the mean grain size increases with increasing $T_s$ for the thick films. The mean grain sizes of the thick films were also measured by a scanning electron microscope. It is found that the difference of the mean sizes determined by the two methods is less than 10\%.

\begin{figure}[htp]
\includegraphics[scale=0.8]{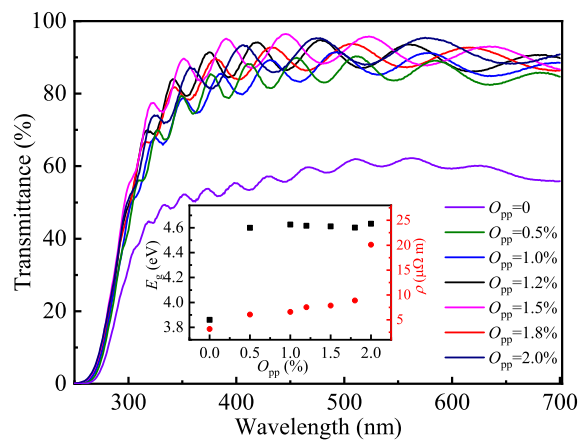}
\caption{(Color online) Room temperature optical transmission spectra of the Ta:SnO$_2$ films deposited at different oxygen partial pressures. Inset: Optical band gap and resistivity versus oxygen partial pressure.}\label{Fig-Trasmitance}
\end{figure}

Figure~\ref{Fig-Trasmitance} shows the optical transmittance ($T_R$) as a function of wavelength for films deposited at $T_s=963$ K and different oxygen partial pressure ($O_{\rm{pp}}$), as indicated. Here $T_R$ is the intensity ratio of transmitted
light to incident light, and the thicknesses of these films are in the range of $\sim$550 to $\sim$1000 nm. For the film deposited in pure argon, the transmittance in the visible range is less than 60\%, while for these deposited at O$_2$ and Ar mixture, the transmittance is greater than 85\%. From Fig.~\ref{Fig-Trasmitance}, one can get the optical absorption coefficient $\alpha$ via $\alpha=-(\ln T_R)t^{-1}$ with $t$ being the thickness of the film. According to Tauc \emph{et al}.,~\cite{tauc_optical_1966} the relation between $\alpha$ and the incident photo energy is $(\alpha h\nu)^{1/n}=A(h\nu-E_g)$, where $n=1/2$ for a direct transition and $n=2$ for an indirect transition, $A$ is a constant, and $E_g$ is the optical band gap. According to the first-principle calculation result,~\cite{darriba_ab_2014} Ta:SnO$_2$ is a direct gap degenerate semiconductor, thus the value of $E_g$ of each film can be obtained by extrapolating the linear part of $(\alpha h\nu)^2$ versus $h\nu$ plot to $(\alpha h\nu)^2=0$. The inset of Fig.\ref{Fig-Trasmitance} shows $E_g$ versus the oxygen partial pressure $O_{\rm{pp}}$ for these Ta:SnO$_2$ films. The optical band gap $E_g$ is $\sim$3.9\,eV for the film deposited at pure argon atmosphere. When 0.5\% oxygen is introduced, $E_g$ increases to $\sim$4.6\,eV and then almost keeps as a constant with further increasing oxygen percentage. The inset of Fig.~\ref{Fig-Trasmitance} also gives the resistivity versus oxygen partial pressure measured at 300\,K. The resistivity increases with increasing  $O_{\rm{pp}}$ in the whole testing range. Combining the oxygen partial pressure dependence of optical transmittance and resistivity, one can see that the optimum oxygen percentage is $O_{\rm{pp}}\simeq 0.5\%$. Thus the optimal depositing conditions for Ta:SnO$_2$ film with both high conductivity and high optical transparency are $T_s \simeq 963$\,K and $O_{\rm{pp}}\simeq 0.5\%$ in our sputtering system.

\begin{table*}
\caption{\label{Table-I} Relevant parameters for some representative Ta:SnO$_2$ films. Here $t$ is the mean film thickness, $T_s$ is the substrate temperature, $d$ is mean diameter of the grains, $\rho_0$ is the residual resistivity, $n^\ast$ is the mean
value of carrier concentrations between 180 and 250\,K, and $\beta_{\rm{BG}}$, $B$, and $\theta_D$ are the adjusting parameters in fitting Eq.~(\ref{Eq-Total}). The parameters $g_T$ is obtained by least-squares fits to Eq.~(\ref{Eq-Cond-gr-R}), and $c_d$ is the adjusting parameter in Eq.~(\ref{Eq-Hall-R}). }
\begin{ruledtabular}
\begin{center}
\begin{tabular}{cccccccccccc}
     &$t$ &$T_s$ & $d$   &$\rho$(300\,K)  & $\rho_0$        & $\beta_{\rm{BG}}$      & $B$                 &$\theta_D$&$n^\ast$    &$g_T$& $c_d$ \\
Films&(nm)& (K)  & (nm)  &($\mu\Omega$\,m)&($\mu\Omega$\,m) &(n$\Omega$\,m\,K$^{-1}$)&(10$^{-7}$\,K$^{-2}$)&  (K)     &  (m$^{-3}$)&     &       \\  \hline
  1  &958 & 983  & 47.3  & 4.46           & 4.23           &  4.81                  & 1.31                & 1024     &            &     &       \\
  2  &992 & 963  & 36.9  &3.16            & 2.90           &  4.07                  & 5.72                & 1116     &            &     &       \\
  3  &962 & 953  & 34.2 & 4.56           & 4.35           &  4.55                  & 0.94                & 1080     &            &     &       \\
  4  &960 & 933  & 26.2  &3.93            & 3.69           &  4.25                  & 2.66                & 1061     &            &     &       \\
  5  & 36 & 963  & 31.6  &8.85            & 8.53           &                        &                     &          & 3.75       &24.02&  2.12 \\
  6  & 30 & 963  & 24.2  &10.93           & 10.64          &                        &                     &          & 3.22       &14.84&  1.46 \\
  7  & 18 & 963  & 17.7  &10.25           & 9.96           &                        &                     &          & 2.95       &10.38&  1.64 \\
  8  & 11 & 963  & 9.8   & 13.04          & 12.9           &                        &                     &          & 3.09       &4.94 &  1.66 \\
\end{tabular}
\end{center}
\end{ruledtabular}
\end{table*}

\subsection{Electron-phonon-impurity interference effect in the thick films}\label{Sub-S2}
\begin{figure}
\includegraphics[scale=0.81]{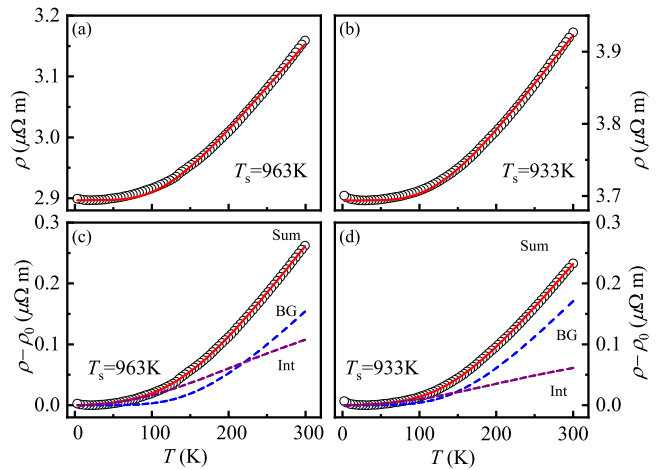}
\caption{(Color online) Temperature dependence of resistivity for the thick  Ta:SnO$_2$  films deposited at 963\,K [(a) and (c)], and 933\,K [(b) and (d)]. The solid curves in (a) and (b) are least-squares fits to Eq.~(\ref{Eq.Bloch}), while those in (c) and (d) are least-squares fits to Eq.~(\ref{Eq-Total}).}\label{Fig-rho-T-thick}
\end{figure}

We firstly investigate the electrical transport properties of the Ta:SnO$_2$  films deposited in pure argon. Figures~\ref{Fig-rho-T-thick}(a) and \ref{Fig-rho-T-thick}(b) show the temperature dependence of resistivity for two representative thick films deposited at $T_s=933$ and 963\,K, respectively. Upon cooling from 300\,K, the resistivity decreases with decreasing temperature, and reaches its minimum at $T_{\rm min}$ (the value of $T_{\rm min}$ depends on the film and is 30\,K and 20\,K for the films deposited at $T_s=933$ and 963\,K, respectively), and then slightly increases with further decreasing temperature. The slight enhancement in resistivity with decreasing temperature could originate from weak-localization (WL) effect and electron-electron interaction (EEI) effect.~ \cite{lee_disordered_1985} The reduction of resistivity with decreasing temperature above $T_{\rm min}$ indicates that Ta:SnO$_2$ film reveals degenerate semiconductor characteristics in electrical transport properties. It is also found that the carrier (conduction electron) concentrations of the thick films are insensitive to temperature from room temperature down to liquid helium temperature, which further confirms the degenerate semiconductor characteristics of the films.

Generally, the resistivity for a degenerate semiconductor or metal comes from two effects: the residual resistivity caused by collisions of electrons with impurity atoms and mechanical imperfections in lattice and the temperature-dependent resistivity originated from electron-phonon scattering. According to Altshuler\cite{altshuler_1978,altshuler_effects_1982,altshuler_1985}, the temperature dependence of resistivity caused by the ``pure'' electron-phonon scattering in disordered metal is similar to the Bloch-Gr\"{u}neisen formula. Thus the total resistivity can be written as\cite{Ziman_1960,Grimvall_1981,ptitsina_electron-phonon_1997,LiZQ_2004}
\begin{equation}\label{Eq.Bloch}
\rho=\rho_0+\beta_{\rm BG}T\left(\frac{T}{\theta_D}\right)^4\int_{0}^{\theta_D/T}\frac{x^5 \mathrm{d}x}{(e^x-1)(1-e^{-x})},
\end{equation}
where $\rho_0$ is the residual resistivity, $\beta_{\rm BG}$ is a constant, and $\theta_D$ is the Debye temperature. The theoretical predictions of Eq.~(\ref{Eq.Bloch}) are least-squares fitted to the $\rho(T)$ data of the Ta:SnO$_2$ films from $T_{\rm min}$  to 300 K  and the results are shown as solid curves in Figs.~\ref{Fig-rho-T-thick}(a) and ~\ref{Fig-rho-T-thick}(b). Clearly, the overall trends of the experimental data can be described by Eq.~({\ref{Eq.Bloch}}). However, a close inspection indicates that the $\rho(T)$ data below $\sim$120 K remarkably deviate from the theoretical predictions. Thus other scattering processes could be neglected in above analysis.

Since 1987, Reizer and Sergeev have realized the importance of the inelastic electron scattering from vibrating impurities in dirty metals.~\cite{Reizer_Sergeev_1987} In this case, the contribution to the resistivity due to the interference between the electron-phonon and electron-impurity scattering can be expressed as\cite{echternach_evidence_1993,yeh_electron-phonon-impurity_2005,ilin_interrelation_1998}
\begin{equation}\label{Eq.Int}
\rho_{\rm int}=BT^2\rho_0\left(\frac{6}{\pi^2}\right)\int_{0}^{\theta_D/T}\left[\frac{x^2e^x}{(e^x-1)^2}-\frac{x}{e^x-1}\right]{\mathrm{d}}x,
\end{equation}
where $B$ is a material-related constant and can be approximately written as  $B\approx 2\pi^2k_B^2u_l^2/(3\epsilon_F p_F u_t^3)$.
Here $k_B$ is the Bolzmann constant, $\epsilon_F$ is Fermi energy, $p_F$ is the Fermi momentum, and $u_l$ and $u_t$ are the sound speed of longitudinal and transverse phonons. For $T\ll \theta_D$, the integral in the right hand side of Eq.~(\ref{Eq.Int}) tends to $\pi^2/6$, then Eq.~(\ref{Eq.Int}) is simplified as  $\rho_{\rm int}\simeq\rho_0BT^2$. Considering the contribution of interference mechanism, one can obtain
\begin{equation}\label{Eq-Total}
  \rho=\rho_0+\rho_{\rm BG}+\rho_{\rm int},
\end{equation}
where $\rho_{\rm BG}$ is the resistivity caused by ``pure" electron-phonon scattering and is represented by the second term in the right hand side of Eq.~(\ref{Eq.Bloch}). The theoretical predictions of Eq.~(\ref{Eq-Total}) are least-squares fitted to the experimental data in Figs.~\ref{Fig-rho-T-thick}(a) and \ref{Fig-rho-T-thick}(b), and the results are redrawn in Fig.~\ref{Fig-rho-T-thick}(c) and \ref{Fig-rho-T-thick}(d), respectively, for clarity. The optimal parameters obtained from the fitting processes are listed in Table~{\ref{Table-I}}. In Fig.~\ref{Fig-rho-T-thick}(c) and ~\ref{Fig-rho-T-thick}(d), we also give the temperature dependence of $\rho_{\rm BG}$ and  $\rho_{\rm int}$ for each film. Clearly, the interference mechanism dominates over  the ``pure" electron-phonon scattering term at low temperature regime (the exact value of the up bound temperature below which $\rho_{\rm BG}$ is less than $\rho_{\rm int}$ is sample dependent). These results indicate that the effect of interference between the electron-phonon and electron-impurity scattering plays a key role in the low-temperature electrical transport processes of Ta:SnO$_{2}$ films. We note in passing that the electron-phonon-impurity interference effect can be neglected in the films deposited at $T_s\simeq 963$\,K and $O_{\rm{pp}}\gtrsim 0.5\%$. The reason needs further investigations.

\begin{figure}[htp]
\includegraphics[scale=0.75]{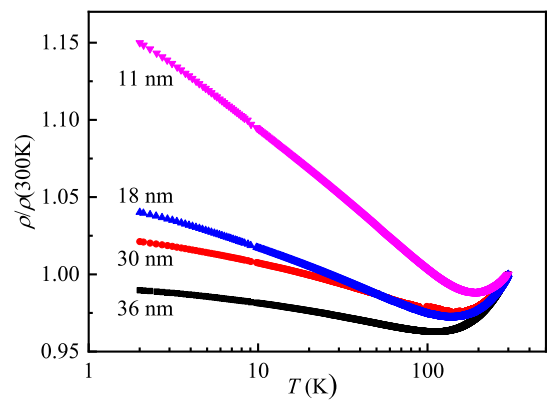}
\caption{(Color online) Normalized resistivity $\rho/\rho(300\,\rm{K})$ varies as a function of temperature for four thin Ta:SnO$_2$  films with different thickness.}\label{Fig-rho-T-Ultrathin}
\end{figure}

\subsection{Intergrain electron-electron interaction effect on the transport in the thin films}
Now we discuss the longitudinal and Hall transport properties of Ta:SnO$_2$ films with thickness ranging from $\sim$11 to $\sim$36 nm. Figure~\ref{Fig-rho-T-Ultrathin} shows the temperature dependence of resistivity of these thin films deposited at $T_s =963$ K and $O_{\rm{pp}}=0.5\%$. Although the overall variation trends of $\rho(T)$ curves for these films are similar to those of the thick Ta:SiO$_2$ films, the enhancement of resistivity with decreasing temperature below $T_{\rm min}$ is more remarkable than that in those thick films. In addition, the resistivity approximately varies linearly with $\log_{10} T$ (or $\ln T$) below $\sim$$T_{\rm min}$. This indicates that the increasing in resistivity with decreasing temperature in the low temperature regime could originate from corrections of WL and EEI effects in two-dimensional (2D) disordered conductors.~\cite{altshuler_magnetoresistance_1980}

The WL and EEI effects in homogeneous disordered metals have been intensively investigated since 1980s.~\cite{altshuler_magnetoresistance_1980,altshuler_interaction_1980,blanter_electron-electron_1998,goh_electron-electron_2008} In 2D homogeneous disordered metals, the Altshuler-Aronov type EEI effect gives the follow correction to the conductivity\cite{altshuler_magnetoresistance_1980,altshuler_interaction_1980}
\begin{equation}\label{A-AEEI}
  \delta\sigma_\square=\frac{e^2}{4 \pi^2 \hbar}(2-2F)\ln(k_BT\tau/\hbar),
\end{equation}
where $F$ is an electron screening factor and $\tau$ is the elastic scattering time. When a magnetic field  is applied, the correction can be expressed as\cite{lee_magnetoresistance_1982,lee_disordered_1985,malshukov_magnetoresistance_1997,bergmann_weak_2010}
\begin{equation}\label{A-AEEIH}
  \delta\sigma_\square=\frac{e^2}{4 \pi^2 \hbar}[(2-2F)\ln(k_BT\tau/\hbar)-G_2(b)],
\end{equation}
where $b=g\mu_BB/k_BT$ with $g$ being the Land\'{e} g-factor, the function $G_2(b)$ can be computed numerically and has the value $\ln(b/1.3)$ and $0.084 b^2$ for $b\gg 1$ and $b\ll 1$, respectively.~\cite{lee_magnetoresistance_1982}

Since 2002, it has been realized that the  EEI effect in granular metals is distinct from that in ``homogeneous disordered metals".~\cite{efetov_transition_2002,beloborodov_transport_2003,Eftov-PRB-2003,beloborodov_granular_2007,kharitonov_hall_2007,kharitonov_hall_2008} The term granular metals refers to composite materials consisting of immiscible metals and insulators. In granular metals, the intergrain electron and intragrain dynamics both play key roles in the electrical transport properties. In strong intergrain coupling regime, [i.e., $1 \ll g_T \ll g_0$, where $g_T =G_T/(2e^2/\hbar)$, $g_0 =G_0/(2e^2/\hbar)$, $G_T$ and $G_0$  are intergrain tunneling conductance and conductance of a metal grain, respectively], the EEI effect in the
presence of granularity gives distinct corrections to both logarithmic conductivity and Hall coefficient. According to the theory, in the temperature range $g_T\delta/k_B \lesssim T \lesssim E_c/k_B$ ($\delta$ is the mean-energy level spacing in the grain and $E_c$ is the charging energy), the conductivity is  determined by the granular structure and incoherent tunneling processes, and can be written as\cite{efetov_transition_2002,beloborodov_transport_2003,Eftov-PRB-2003,beloborodov_granular_2007}
\begin{equation}\label{Eq.(conductance-graular)}
\sigma(T)=\sigma_{0}\left[1-\frac{1}{2\pi g_T \tilde{d}}\ln\left(\frac{g_{T}E_{c}}{k_{B}T}\right)\right],
\end{equation}
where $\sigma_{0}$ is the conductivity without the EEI effect and $\tilde{d}$ is the dimensionality of the granular array. In the temperature range $g_T\delta/k_B \lesssim T \lesssim \min(g_T E_c, E_{\rm Th})/k_B$, the virtual electron diffusion inside individual grains governs the temperature dependent behavior of the Hall coefficient\cite{kharitonov_hall_2007,kharitonov_hall_2008}
\begin{equation}\label{Eq.(Hall)}
R_H=\frac{1}{n^\ast e} \left[ 1+\frac{c_{d}}{4\pi g_T} \ln \left(\frac{\min(g_{T}E_{c}, E_{\text{Th}})}{k_{B}T}\right) \right],
\end{equation}
where $n^\ast$ is the effective carrier concentration, $c_d$ is a numerical lattice factor of order unity, $E_{\text{Th}}$ is the Thouless energy. It should be noted  that the $\ln T$ dependent behaviors of $\sigma$ in Eq.~(\ref{Eq.(conductance-graular)}) and $R_H$ in Eq.~(\ref{Eq.(Hall)}) are independent of the granular array dimensionality, and are the consequence of and specific to the granularity.

\begin{figure}
\includegraphics[scale=0.75]{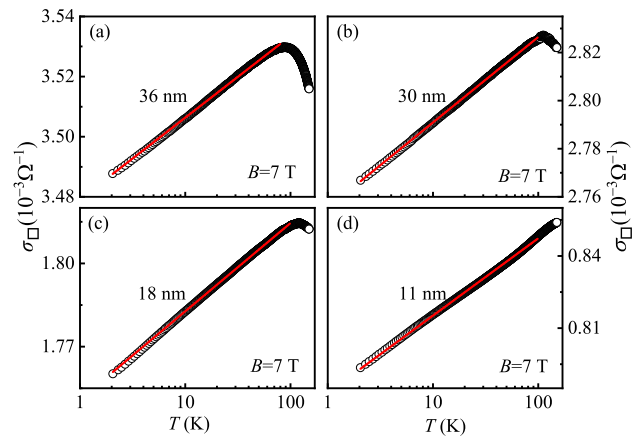}
\caption{(Color online) Variation in sheet conductivity with a logarithm of temperature for four thin Ta:SnO$_2$ films [(a) $t= 36$\,nm, (b) $t=30$\,nm, (c) $t=18$\,nm, and (d) $t=11$\,nm] measured
in a perpendicular magnetic field of 7\,T. The solid straight lines are
least-squares fits to Eq.~(\ref{Eq-Cond-gr-R}).}\label{Fig-sigma-Ultrathin}
\end{figure}

To understand which kind of EEI effect governs the low temperature transport properties of these thin films, we measured the temperature dependence of conductivity of the films in a perpendicular magnetic field $B$ of 7\,T, which is sufficiently large to suppress the WL effect.~\cite{B_phi} Figure~\ref{Fig-sigma-Ultrathin} shows the sheet conductivity $\sigma_\square$ as a function of logarithm of temperature, where $\sigma_\square$ is the reciprocal of the sheet resistance. Clearly, the sheet conductivity varies linearly with $\log_{10} T$, or $\ln T$ from 2 to $\sim$90\,K. We firstly compare the $\sigma_\square(T)$ data with Eq.~(\ref{A-AEEIH}). At low temperature and large magnetic field, Eq.~(\ref{A-AEEIH}) can be rewritten as
\begin{equation}\label{A-AEEIH-1}
 \sigma_\square(T)=\sigma_\square(T_0)+\frac{e^2}{4\pi^2\hbar}(2-F)\ln\left(\frac{T}{T_0}\right),
\end{equation}
where $T_0$ is an arbitrary reference temperature. Taking $T_0=2$\,K, we fit the $\sigma_\square(T)$ data to Eq.~(\ref{A-AEEIH-1})using least-square method.  Although the experimental $\sigma_\square (T)$ data overlap with the theoretical predictions of Eq.~(\ref{A-AEEI}), the fitted values of $F$ are all negative, e.g. $F\simeq -0.29$, $-0.47$, $-0.28$, and $-0.21$ for the $t\simeq 36$, 30, 18, and 11\,nm  films, respectively. Theoretically, the value of the screening factor $F$ should be $0\lesssim F \lesssim 1$. Thus the negative value of $F$ indicates that the $\ln T$ behavior of $\sigma_\square (T)$ does not originate from the Altshuler-Aronov type EEI effect.

Next we analyze the $\sigma_\square (T)$ data in the framework of the EEI theory in the presence of granularity. From Table~{\ref{Table-I}}, one can see that the  mean grain sizes of the $t\simeq 36$, 30, 18, and 11\,nm films  are 31.6, 24.2, 17.7 and 9.8\,nm, respectively, thus these films are nominally covered with only one layer of Ta:SnO$_2$ grains and the dimensionality of the grain array is $\tilde{d}=2$ for each film. For 2D granular system, we rewrite Eq.~(\ref{Eq.(conductance-graular)}) as
\begin{equation}\label{Eq-Cond-gr-R}
  \sigma_\square(T)=\sigma_\square(T_0)+\frac{\sigma_{\square}^{0}}{4\pi g_T}\ln\left(\frac{T}{T_0}\right),
\end{equation}
where $\sigma_\square^0$ is the sheet conductance [corresponding to the $\sigma_0$ in Eq.~(\ref{Eq.(conductance-graular)})] without the EEI effect.
Treating $g_T$ as an adjusting parameter, we fit the $\sigma_\square(T)$ data to Eq.~(\ref{Eq-Cond-gr-R}) by taking $T_0=2$\,K and $\sigma_\square^0=\sigma(90\,\rm{K})$ and the results are plotted as solid lines in Fig.~\ref{Fig-sigma-Ultrathin}. Inspection of Fig.~\ref{Fig-sigma-Ultrathin} indicates that the $\sigma_\square(T)$ data for the four films can be well described by Eq.~(\ref{Eq-Cond-gr-R}). In addition, the fitted values of the adjustable parameter $g_T$ all satisfy, $g_T\gg 1$ (see Table~{\ref{Table-I}}), the prerequisite for the validation of Eq.~(\ref{Eq-Cond-gr-R}).

\begin{figure}[htp]
\includegraphics[scale=0.75]{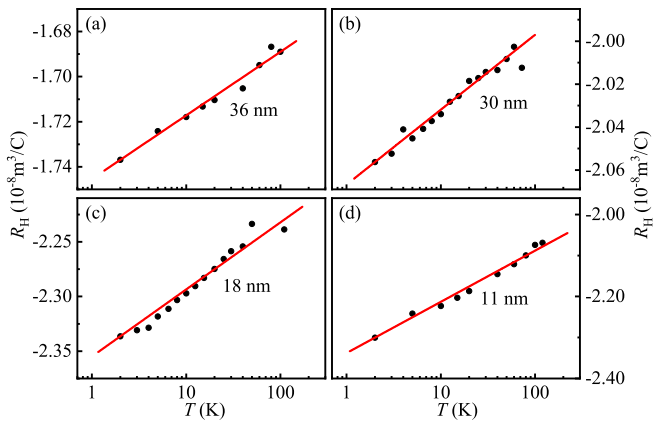}
\caption{(Color online) Hall coefficient as a function of logarithm of temperature for four thin Ta:SnO$_2$ films with (a) $t= 36$\,nm, (b) $t=30$\,nm, (c) $t=18$\,nm, and (d) $t=11$\,nm. The solid straight lines are least-squares fits to Eq.~(\ref{Eq-Hall-R}).}\label{FigHall}
\end{figure}

Figure~\ref{FigHall} shows the Hall coefficient varies as a function of temperature for the four thin films, where the temperature axis is plotted in logarithmic scale.
Clearly, the Hall coefficients vary linearly with $\ln T$. Similarly, Eq.~(\ref{Eq.(Hall)}) is rewritten as
\begin{equation}\label{Eq-Hall-R}
  R_H(T)=R_H(T_0)+\frac{1}{n^\ast e}\frac{c_d}{4\pi g_T}\ln\left(\frac{T_0}{T}\right),
\end{equation}
Our measured $R_H(T)$ data are least-squares fitted to Eq.~(\ref{Eq-Hall-R}) and the fitted results are represented by the solid lines in Fig.~\ref{FigHall}. In the fitting processes, $c_d$ is the adjustable parameter, $g_T$ has been obtained in fitting the $\sigma_\square(T)$ data, $T_0$ is set as 2\,K, and $n^\ast$ is taken the mean value measured between 180 and 250 K, in which the EEI correction to $R_H$ can be neglected. The values of $c_d$ and $n^\ast$ for each film are listed in Table~\ref{Table-I}. Clearly, the $R_H(T)$ data of these thin films are overlapped with the theoretical prediction of Eq.~(\ref{Eq.(Hall)}). The values of $c_d$ are in the reasonable range. Thus Eqs.~(\ref{Eq-Cond-gr-R}) and (\ref{Eq-Hall-R}) are safely applicable for the thin films.

Before closing this subsection, we summarize the experimental advances in the study of the influence of Coulomb effects on the transport properties in granular metals. Soon after the theoretical finding of the $\delta\sigma\propto\ln T$ law [Eq.~(\ref{Eq.(conductance-graular)})], it was tested in several granular systems, including Pt/C composite nanwires,~\cite{Rotkina-PRB2005,Sachser-PRL2011} B-doped diamond films,~\cite{Achatz-PRB2009} and granular Cr films.~\cite{Sun-PRB2010} Thus the $\delta\sigma\propto\ln T$ law was verified. To verify $\delta R_H \propto\ln T$ law [Eq.~(\ref{Eq.(Hall)})], one generally needs to measure the Hall voltage versus magnetic field at each testing temperature. This is much more difficult than the conductivity versus temperature measurement (the logarithmic correction term $\delta R_H$ is estimated to be less than $\sim$10\% of the total $R_H$).
Taking advantage of the low carrier concentration and degenerate semiconductor characteristics of tin-doped indium oxide (ITO) films, Zhang \emph{et al}.~\cite{zhangYJ2011PRB} verified Eq.~(\ref{Eq.(Hall)}) as well as Eq.~(\ref{Eq.(conductance-graular)}) in 2D ITO granular films. Shortly afterwards, Yang \emph{et al}.~\cite{yang_influence_2012} also demonstrated the validity of Eq.~(\ref{Eq.(Hall)}) and Eq.~(\ref{Eq.(conductance-graular)}) in 2D Al-doped ZnO granular arrays. In 2015, Wu \emph{et al}. measured the temperature dependence of conductivities and Hall coefficients of in  2D and 3D Ag$_x$(SnO$_2$)$_{1-x}$ ($x$ being the volume fraction of Ag) granular composites.~\cite{wu_electron-electron_2015} They observed a $\delta R_H \propto\ln T$ law as well as a $\delta\sigma\propto\ln T$ law in a wide temperature range. These $\ln T$ temperature behaviors are independent of the dimensionality of granular arrays, and can be explained by the EEI effects in the presence of granularity. In the thin Ta:SnO$_2$ films (this paper), each grain is treated as a metallic particle ($g_T\ll g_0$) and the carrier concentration in each film is $\sim$2 to 3 orders of magnitude lower than those in typical metals (which facilitates the measurement of the temperature dependence of $R_H$). The experimental $\sigma (T)$ [$R_H (T)$] data are least-square fitted to Eq.~(\ref{Eq-Cond-gr-R}) [Eq.~(\ref{Eq-Hall-R})] instead of Eq.~(\ref{Eq.(conductance-graular)}) [Eq.~(\ref{Eq.(Hall)})]. This ensures only one parameter [$g_T$ in Eq.~(\ref{Eq-Cond-gr-R}) or $c_d$ in Eq.~(\ref{Eq-Hall-R})] can be adjusted in the fitting process. Thus the uncertainty arisen in estimating $E_c$ or $E_{\rm Th}$ is avoided. In fact, the value of $E_c$ or $E_{\rm Th}$ is difficult to be estimated in 2D granular arrays with disk-shaped granules.\cite{Sun-PRB2010,zhangYJ2011PRB,yang_influence_2012} Overall, our results in the thin Ta:SnO$_2$ films further confirm that the EEI effects on the transport properties in granular metals are different from that in homogeneous disordered metals.

\section{Conclusion}
In summary, Ta:SnO$_2$ films with low resistivity and high optical transparency are successfully fabricated by rf-sputtering methods. It is found that the optimal depositing conditions for low resistivity and high optical transparency are $T_s \simeq 963$\,K and $O_{\rm pp}\simeq 0.5\%$. The electrical transport properties of the thick films and thin films are systematically investigated. It is found that all films have degenerate semiconductor characteristics. For the thick films deposited at pure argon, besides the electron-phonon scattering, the interference effect between electron-phonon and electron-impurity scattering also plays an important role in determining the temperature dependent behaviors of the resistivity. For the thin films, the robust $\Delta \sigma_\square\propto\ln T$ law and $\Delta R_H \propto \ln T$ law have been observed from liquid helium temperature to $\sim$100\,K. We have found that the $\ln T$ behaviors of the conductivity and Hall coefficient originate from the EEI effect in the presence of granularity. Our results show that the Ta:SnO$_2$ film, although a new TCO material, can be a model system for exploration new electrical transport phenomena.

\begin{acknowledgments}
This work is supported by the National Natural Science Foundation of China through Grant No. 11774253 and 12174282.
\end{acknowledgments}

\section*{DATA AVAILABILITY}
The data that support the findings of this study are available from the corresponding author upon reasonable request.


\begin{thebibliography}{00}\label{sec:TeXbooks}
\bibitem{Jarzebski1982}Z. M. Jarzebski, Phys. Stat. Sol. \textbf{71}, 13 (1982).
\bibitem{exarhos_discovery-based_2007} G. J. Exarhos and X. D. Zhou, Thin Solid Films \textbf{515}, 7025 (2007).
\bibitem{David2010}S. G. David and D. Clark, MRS. Bulletin. \textbf{25}, 15 (2010).
\bibitem{Gilshtein2020}E. Gilshtein, S. Bolat, G. T. Sevilla, A. C. Vidani, Clemens, T. Graule, A. N. Tiwari, and Y. E. Romanyuk, Adv. Mater. Technol. \textbf{5}, 2000369 (2020).
\bibitem{Singh2008} A. K. Singh, A. Janotti, M. Scheffler, and C. G. V. Walle, Phys. Rev. Lett. \textbf{101}, 055502 (2008).
\bibitem{Kim2008}H. Kim, G. P. Kushto, R. C. Y. Auyeung, and A. Piqu\'{e}, Appl. Phys. A \textbf{93}, 521 (2008).
\bibitem{Hu2011}Z. Hu, J. Zhang, Z. Hao, Q. Hao, X. Geng, and Y. Zhao, Appl. Phys. Lett. \textbf{98}, 123302 (2011).
\bibitem{Gordon2000}R. G. Gordon, MRS. Bulletin. \textbf{25}, 52 (2000).
\bibitem{king_conductivity_2011}P. D. C. King and T. D. Veal, J. Phys.: Conden. Matter \textbf{23}, 334214 (2011).
\bibitem{gao_solid-state_2012}B. H. Gao, S. N. Ding, Q. Q. Li, D. S. Shan, M. Yue, and S. Cosnier, Electroanal. \textbf{24}, 1267 (2012).
\bibitem{islam_alkaline_2020} M. A. Islam, J. R. Mou, M. F. Hossain, A. M. M. T. Karim, M. Kamruzzaman, and M. S. Hossain, J. Solgel Sci. Technol. \textbf{96}, 304 (2020).
\bibitem{rakhshani_electronic_1998}A. E. Rakhshani, Y. Makdisi, and H. A. Ramazaniyan, J. Appl. Phys. \textbf{83}, 1049 (1998).
\bibitem{shi_roughness_2014}X. L. Shi, J. T. Wang, and J. N. Wang, J. Alloys Compd. \textbf{611}, 297 (2014).
 \bibitem{suh_atmospheric-pressure_1999}S. Suh, Z. H. Zhang, W. K. Chu, and M. H. David, Thin Solid Films \textbf{345}, 240 (1999).
\bibitem{lee_structural_2006}S. Y. Lee and B. O. Park, Thin Solid Films \textbf{510}, 154 (2006).
\bibitem{kim_doping_2016}C. Y. Kim, E. B. Go, J. S. Choi, and S. C. Choi, J. Nanosci. Nanotechnol. \textbf{16}, 11330 (2016).
\bibitem{jain_electrical_2004}G. Jain and R. Kumar, Opt. Mater. \textbf{26}, 27 (2004).
\bibitem{kim_transparent_2008}H. Kim, R. C. Y. Auyeung, and A. Piqu\'{e}, Thin Solid Films \textbf{516}, 5052 (2008).
\bibitem{schleife_ab_2012}A. Schleife and F. Bechstedt, J. Mater. Res. \textbf{27}, 2180 (2012).
\bibitem{darriba_ab_2014}Germ\'{a}n N. Darriba, Emiliano L. Mu\~{n}oz, Leonardo A. Errico, and Mario. Renter\'{\i}a, J. Phys. Chem. C. \textbf{118}, 19929 (2014).
\bibitem{behtash_electronic_2015}M. Behtash, P. H. Joo, S. Nazir, and K. S. Yang, J. Appl. Phys. \textbf{117}, 175101 (2015).
\bibitem{birks_particle_1946}L. S. Birks and H. Friedman, J. Appl. Phys. \textbf{17}, 687 (1946).
\bibitem{ishikawa_size_1988}K. Ishikawa, K. Yoshikawa, and N. Okada, Phys. Rev. B \textbf{37}, 5852 (1988).
 \bibitem{yang_nanoscaled_2002}X.C. Yang, W. Riehemann, M. Dubiel, and H. Hofmeister, Mater. Sci. Eng. B \textbf{95}, 299 (2002).
\bibitem{mariappan_synthesis_2006}C.R. Mariappan, C. Galven, M.P. Crosnier-Lopez, F. Le Berre, and O. Bohnke, J. Solid State Chem. \textbf{179}, 450 (2006).
\bibitem{tauc_optical_1966}J. Tauc, R. Grigorovici, and A. Vancu, Phys. Status Solidi \textbf{15}, 627 (1966).
\bibitem{lee_disordered_1985}P. A. Lee and T. V. Ramakrishnan, Rev. Mod. Phys. \textbf{57}, 287 (1985).
\bibitem{altshuler_1978}B. L. Altshuler, Zh Eksp. Teor. Fiz. \textbf{75}, 1330 (1978) [Sov. Phys. JEPT \textbf{48}, 670 (1978)].
\bibitem{altshuler_effects_1982}B. L. Altshuler, A. G. Aronov, and D. E. Khmelnitsky, J. Phys. C: Solid State \textbf{15}, 7367 (1982).
\bibitem{altshuler_1985}B. L. Altshuler and A. G. Aronov, in \emph{Electron-Electron Interactions in Disordered Systems}, edited by A. L. Efros and M. Pollak (Elsevier, Amsterdam, 1985).
\bibitem{Ziman_1960} J. M. Ziman, \emph{Electron and Phonons} (Clarendon Press, Oxford, 1960).
\bibitem{Grimvall_1981}G. Grimvall, \emph{The Electron-Phoneon Interaction in Metal} (North-Holland, Amsterdam, 1981).
\bibitem{ptitsina_electron-phonon_1997}N. G. Ptitsina, G. M. Chulkova, K. S. Il’in, A. V. Sergeev, F. S. Pochinkov, E. M. Gershenzon, and M. E. Gershenson, Phys. Rev. B \textbf{56}, 10089 (1997).
\bibitem{LiZQ_2004} Z. Q. Li and J. J. Lin, J. Appl. Phys. \textbf{96}, 5918 (2004)
\bibitem{Reizer_Sergeev_1987}M. Yu. Reizer and A. V. Sergeev, Zh. Eksp. Teor. Fiz. \textbf{92}, 2291 (1987) [Sov. Phys. JETP \textbf{65}, 1291 (1987)].
\bibitem{echternach_evidence_1993}P. M. Echternach, M. E. Gershenson, and H. M. Bozler, Phys. Rev. B \textbf{47}, 13659 (1993).
\bibitem{yeh_electron-phonon-impurity_2005}S. S. Yeh, J. J. Lin, X. N. Jing, and D. L. Zhang, Phys. Rev. B \textbf{72}, 024204 (2005).
\bibitem{ilin_interrelation_1998} K. S. Il’in, N. G. Ptitsina, A. V. Sergeev, G. N. Gol’tsman, E. M. Gershenzon, B. S. Karasik, and E. V. Pechen, Phys. Rev. B \textbf{57}, 15623 (1998).
\bibitem{altshuler_magnetoresistance_1980}B. L. Altshuler, D. Khmel'nitzkii, A. I. Larkin, and P. A. Lee, Phys. Rev. B \textbf{22}, 5142 (1980).
\bibitem{altshuler_interaction_1980}B. L. Altshuler, A. G. Aronov, and P. A. Lee, Phys. Rev. Lett. \textbf{44}, 1288 (1980).
\bibitem{blanter_electron-electron_1998}Y. M. Blanter and A. D. Mirlin, Phys. Rev. B \textbf{57}, 4566 (1998).
\bibitem{goh_electron-electron_2008}K. E. J. Goh, M. Y. Simmons, and A. R. Hamilton, Phys. Rev. B \textbf{77}, 235410 (2008).
\bibitem{lee_magnetoresistance_1982}P. A. Lee and T. V. Ramakrishnan, Phys. Rev. B \textbf{26}, 4009 (1982).
 \bibitem{malshukov_magnetoresistance_1997}A. G. Mal’shukov, K. A. Chao, and M. Willander, Phys. Rev. B \textbf{56}, 6436 (1997).
\bibitem{bergmann_weak_2010}G. Bergmann, Int. J. Mod. Phys. B \textbf{24}, 2015 (2010).
\bibitem{B_phi} The characteristic dephasing field for WL $B_\varphi= \hbar/(4eL^{2}_{\varphi})$ with $L_\varphi$ being the dephasing length. $B_\varphi$ is appromately equal to 3.0\,T at 80\,K for the $t=11$\,nm Ta:SnO$_2$ films, and decreases with decreasing temperature. At a certain temperature, the values of $B$ for other thin films are less than that for the 11\,nm thick film.
\bibitem{efetov_transition_2002}K. B. Efetov and A. Tschersich, Europhys. Lett. \textbf{59}, 114 (2002).
\bibitem{beloborodov_transport_2003}I. S. Beloborodov, K. B. Efetov, A. V. Lopatin, and V. M. Vinokur, Phys. Rev. Lett. \textbf{91}, 246801 (2003).
\bibitem{Eftov-PRB-2003}K. B. Efetov and A. Tschersich, Phys. Rev. B \textbf{67}, 174205 (2003).
\bibitem{beloborodov_granular_2007}I. S. Beloborodov, A. V. Lopatin, V. M. Vinokur, and K. B. Efetov, Rev. Mod. Phys. \textbf{79}, 469 (2007).
\bibitem{kharitonov_hall_2007} M. Y. Kharitonov and K. B. Efetov, Phys. Rev. Lett. \textbf{99}, 056803 (2007).
\bibitem{kharitonov_hall_2008}M. Y. Kharitonov and K. B. Efetov, Phys. Rev. B \textbf{77}, 045116 (2008).
\bibitem{Rotkina-PRB2005} L. Rotkina, S. Oh, J. N. Eckstein, and S. V. Rotkin, Phys. Rev. B \textbf{72}, 233407 (2005).
\bibitem{Sachser-PRL2011} R. Sachser, F. Porrati, C. H. Schwalb, and M. Huth, Phys. Rev. Lett. \textbf{107}, 206803 (2011).
\bibitem{Achatz-PRB2009} P. Achatz, W. Gajewski, E. Bustarret, C. Marcenat, R. Piquerel, C. Chapelier, T. Dubouchet, O. A. Williams, K. Haenen, J. A. Garrido, and M. Stutzmann, Phys. Rev. B \textbf{79}, 201203(R) (2009).
\bibitem{Sun-PRB2010} Y. C. Sun, S. S. Yeh, and J. J. Lin, Phys. Rev. B \textbf{82}, 054203 (2010).
\bibitem{zhangYJ2011PRB}Y. J. Zhang, Z. Q. Li, and J. J. Lin, Phys. Rev. B \textbf{84}, 052202 (2011).
\bibitem{yang_influence_2012}Y. Yang, Y. J. Zhang, X. D. Liu, and Z. Q. Li, Appl. Phys. Lett. \textbf{100}, 262101 (2012).
\bibitem{wu_electron-electron_2015}Y. N.Wu, Y. F. Wei, Z. Q. Li, and J. J. Lin, Phys. Rev. B \textbf{91}, 104201 (2015).

\end{thebibliography}
\end{document}